\documentclass{article}
\usepackage{graphicx}

\begin{document}

\title{COMPACT GAMMA DETECTOR MODULE FOR RADIATION MONITORING.}
\author{A. Mammadli, Z. Ahmadov, K. Isayev, N. Sadigova }
\date{{\large Institute of Radiation Problems, Baku, Azerbaijan} \\
  01 August 2022}
\maketitle

\begin{abstract}
Modern unmanned technologies make it possible to implement them in almost all industries, including radiation monitoring. Already in 2011, at the Fukushima nuclear power plant accident, the first drones were demonstrated to measure the radiation background of the territories. Improvement of such devices is one of the topical directions. Our research team has proposed the concept of a gamma radiation detector module for unmanned aerial vehicles. The concept is represented by a matrix of modernized silicon photomultipliers based on silicon structures. The new structures have a design of deep pixel photodiodes of the MLPD type. Today's experimental samples have superiority in such parameters as high efficiency of registration of ionizing radiation. Also, such parameters as low sensitivity to vibrations and low power consumption of photodiodes make it possible to integrate them into unmanned and small-sized radiation monitoring systems.
\end{abstract}

\section{Introduction}

Recently, unmanned aerial vehicles (UAVs) have been increasingly used to obtain operational and reliable information in critical infrastructures\cite{aa}. The advantages of using UAVs are high efficiency, which is especially important in an emergency; exclusion or significant reduction of the threat to the life and health of personnel; visual inspection of objects in real time and ensuring high quality images in difficult post-accident conditions; Possibility to measure emissions and releases of hazardous substances in hard-to-reach areas of facilities. UAVs increasingly being used in the nuclear industry to improve the efficiency of monitoring both before and after an accident. For example, Japanese experts proposed a NPP monitoring system, which includes UAVs (airplanes, helicopters, drones) equipped with video cameras and radiation reconnaissance and dosimetric control devices, a manned helicopter, and other devices. UAVs, to determine the quality of decontamination work at the stage of liquidation of the consequences of an accident, programmed to periodically monitor the radiation situation over the same NPP facilities, having the function of hovering over them and the ability to transmit the collected data in real time to the crisis center.

Radiological monitoring of post-war actions on the territory of Azerbaijan is one of the priority areas in the process of returning Azerbaijani refugees, which called the "big return". However, there are some problems associated with radiation monitoring of territories. These problems include lack of staff, lack of specialized equipment and long monitoring times. There is also a great risk for employees working in the prefrontal area, where there are mined areas.

The high efficiency and efficiency of UAVs, as well as the problems associated with radiological monitoring of territories liberated from occupation, make it relevant in the development of unmanned monitoring systems. However, there are some disadvantages of gamma radiation detectors. These shortcomings are associated with detectors based on photomultiplier tubes, their large size, weight, high power consumption and increased sensitivity to vibrations, etc. makes it ineffective for operation on UAVs. The concept of a new detector of gamma radiation based on silicon photomultiplier described in works\cite{ab}.

Micro pixel avalanche photodiodes, are proposed and studied in \cite{ae}. Our group developed the photodiodes as part of scientific international collaboration activities. Photodiodes differ from their vacuum counterparts in high photon detection efficiency, low operating voltage, small size and high vibration resistance, which makes them prime candidates for unmanned monitoring systems.

\section{Detector and information transmission module.}
 
An MAPD-3NK photodiode and LFS scintillator were used in the development of a scintillation detector for CT\cite{akbarov2018scintillation}. Photodiode MAPD has an area of 3.7 × 3.7 mm, operating voltage 91V, gain 100,000 and photon detection efficiency (PDE) 40\%\cite{akbarov2020scintillation}. The LFS scintillator had dimensions 3x3x10 mm \cite{ahmadov2022investigation}. These types of gamma detector modules are capable of detecting gamma radiation in the energy range from 100 keV to 3 MeV. On Fig.1 shows a block diagram of the detection module (dosimeter)\cite{sadygov2014model}.

A voltage converter circuit was used to supply voltage to the detector. To provide a constant output voltage, a DC voltage converter was assembled, which is controlled by a microcontroller unit. In addition, you can get a constant voltage of +94V at the output by applying a different input voltage (3.5-5V or 22-25.2V) depending on where the detector module will be used. The second part of the detector module is the signal amplifier block. For gamma rays of 60 keV, the amplitude of the signals received at the output of the micropixel photodiode is very small - 10 mV, so such signals are difficult to process. For this reason, their amplitudes enter the analyzer after a 60-fold amplification in the amplifier. The LH6655 operational amplifiers were used for amplification of signals. The third part of the detector module is the analyzer block. The main part of the analyzer unit is an analog-to-digital converter (ADC) and an amplitude analyzer, which is one of the main devices of nuclear electronics and serves to maintain the maximum values of input signals. The microcontroller block processes the signals from the amplitude analyzer and sends them over long distances via the RF module. Depending on the terrain and the distance over which information is transmitted, you can change the power and frequency of the RF module. During the initial checks, the information was transmitted over a distance of 3-5 km and checked. This allows you to monitor an area with a radius of 3-5 km from a safe position. The transmitted information is received by the RX remote control and shown on the display. In addition, it is recorded along with the GPS data from the drone. This information will be used in future area mapping.  

\begin{figure}[ht]
    \centering
    \includegraphics[width=0.6\linewidth]{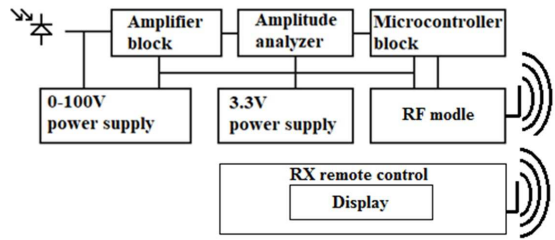} 
    \caption{Scheme of the detection and information transmission module}
    \label{fig:my_label}
\end{figure}

\section{Performance of detection module.}

The detection characteristics of detectors based on the MAPD photodiode and the LFS scintillator were studied. The dimensions of the LFS scintillator are 3 × 3 × 10 mm. The surface of the scintillator was wrapped with several layers of Teflon 100 µm thick. The MAPD photodiode had a dark current of 600 nA, a gain of 40000, a PDE of 40\%, and an operating voltage of 91 V. Am-241, Ba-133 and Na-22 radioisotopes were used as gamma rays. On fig. 2 shows the spectrum of the amplitude distribution of gamma rays emitted by the radioisotope Am-241. The isotope Am-241 is known to emit alpha particles with energies of 5.49 MeV (probability 48.5\%) and 5.44 MeV (probability 13\%), 26.3 keV (probability 2.4\%) and 59.6 keV (probability 2.4\%). with a probability of emitting 35.9\%) also emits energetic gamma rays. As can be seen from the figure, two peaks are observed in the spectrum. The peak in the channel in 2012 belongs to gamma radiation with an energy of 59.6 keV emitted by the Am-241 isotope. The energy resolution of gamma radiation with an energy of 59.6 keV emitted by the isotope Am-241 was 35\%.

\begin{figure}[ht]
    \centering
    \includegraphics[width=0.6\linewidth]{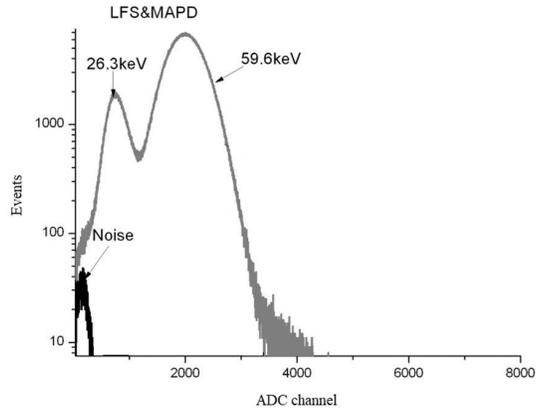}
    \caption{The spectrum of the amplitude distribution of gamma rays emitted by the radioisotope Am-241}
    \label{fig:my_label1}
\end{figure} 

The peak energy per spectrum channel 740 corresponds to 26.3 keV gamma radiation. To verify that this peak is a low-energy gamma-quantum, a thin copper plate 0.2–1 mm thick was placed between the source and the scintillator. In this case, gamma rays entering channel 740 are completely absorbed only in the thickness of 1 mm of the copper layer. This once again confirms that this peak really belongs to a low-energy gamma-ray. The part of the channel that falls in the region 0-350 is the noise generated directly by the dark current of photodetectors of the MAPD type and the electronic circuit. As can be seen in Figure 3, this channel area is observed regardless of the presence of a 1 mm thick copper plate. On fig. 3(a) shows the amplitude spectrum of the Ba-133 and Am-241 radioisotopes (drawn with a 1 mm thick copper plate). It is known that the source of Ba-133 is gamma rays with energies of 81 keV (ejection probability - 31.06\%), 303 keV (probability 18.33\%), 356 keV (probability - 62.05\%), and also emits a characteristic x-ray radiation with an energy of 31 keV. As can be seen from the spectrum, the maximum peak corresponding to a gamma-ray photon with an energy of 31 keV is observed in channel 921. Events corresponding to other gamma rays 81 keV, 303 keV and 356 keV are observed in channels 2644, 5692, 10552, 12557. The reason why the peaks in the low energy region (31-81 keV) are perfectly clear is due to the fact the appearance of Compton scattering in this energy range is very small. The high probability of Compton scattering in high-energy (301-356 keV) gamma quanta did not make it possible to clearly distinguish photopeaks. On fig. 3(b) shows the spectrum of gamma radiation emitted by the Na-22 source. In this case, an additional attenuator was used (1.8 times). An angilation peak corresponding to 511 keV gamma radiation was observed in channel 10211. A peak corresponding to a sum of 1.27 MeV gamma quanta was observed in channel 25685. The Compton edge corresponding to both peaks was observed in full.

\begin{figure}[htb]
    \centering
    \includegraphics[width=0.99\linewidth]{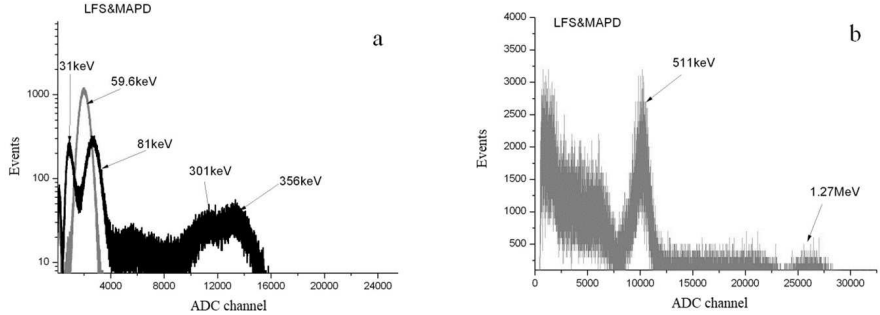}
    \caption{Amplitude distribution of signals detected by Ba-133, Am-241(a) and Na-22(b) radioisotopes based on LFS scintillator + MAPD detector}
    \label{fig:my_label2}
\end{figure} 

\section{Results}

The characteristics of gamma-ray detection by detectors based on MAPD have been studied. It is shown that detectors based on LFS scintillators can distinguish gamma rays in the energy range 26.3 keV-1.33 MeV in amplitude. At the same time, drones can make safer measurements by transmitting information over long distances.

\bibliographystyle{plain}
\bibliography{references.bib}
\end{document}